\documentclass[twocolumn,showpacs,prx,superscriptaddress,floatfix]{revtex4}

\usepackage{color}
\usepackage{tabularx}
\usepackage{epsfig}
\usepackage{amsmath}
\usepackage{amssymb}
\usepackage{graphicx}
\usepackage{wasysym}
\usepackage{dcolumn}
\usepackage{bm}
\usepackage{subfigure}
\usepackage{psfrag}
\usepackage{subfigure}
\usepackage{epstopdf}

\newcommand{\LHx}{{{\rm LiHo_xY_{1-x}F_4}}}
\newcommand{\Ho}{{\rm Ho^+}}

\newcommand{\sss}{\scriptscriptstyle}
\newcommand{\braket}[1]{\left\langle{#1}\right\rangle}
\newcommand{\av}[1]{\left[{#1}\right]_{\sss\rm av}}
\newcommand{\aT}[1]{\braket{#1}_{\sss\rm T}}

\begin{document}

\title{Existence of a Thermodynamic Spin-Glass Phase in the\\
Zero-Concentration Limit of Anisotropic Dipolar Systems}

\author{Juan Carlos Andresen}
\affiliation{Theoretische Physik, ETH Zurich, CH-8093 Zurich,
Switzerland}
\affiliation{Department of Theoretical Physics, Royal Institute of Technology,
SE-106 91 Stockholm, Sweden}

\author{Helmut G.~Katzgraber$^{\star}$}
\affiliation{Department of Physics and Astronomy, Texas A\&M
University, College Station, Texas 77843-4242, USA}
\affiliation{Materials Science and Engineering Program, Texas A\&M
University, College Station, Texas 77843, USA}
\affiliation{Santa Fe Institute, 1399 Hyde Park Road, Santa Fe, New Mexico
87501}

\author{Vadim Oganesyan}
\affiliation{Department of Engineering Science and Physics, College of
Staten Island, CUNY, Staten Island, New York 10314, USA}
\affiliation{Initiative for the Theoretical Sciences, The Graduate
Center, CUNY, New York, New York 10016, USA}

\author{Moshe Schechter}
\affiliation{Department of Physics, Ben Gurion University of the Negev,
Beer Sheva 84105, Israel}

\begin{abstract}

The nature of ordering in dilute dipolar interacting systems dates back
to the work of Debye and is one of the most basic, oldest and as-of-yet
unsettled problems in magnetism. While spin-glass order is readily
observed in several RKKY-interacting systems, dipolar spin-glasses are
the subject of controversy and ongoing scrutiny, e.g., in $\LHx$, a
rare-earth randomly diluted uniaxial (Ising) dipolar system.  In
particular, it is unclear if the spin-glass phase in these paradigmatic
materials persists in the limit of zero concentration or not.  We study
an effective model of $\LHx$ using large-scale Monte Carlo simulations
that combine parallel tempering with a special cluster algorithm
tailored to overcome the numerical difficulties that occur at extreme
dilutions. We find a paramagnetic to spin-glass phase transition for all
$\Ho$ ion concentrations down to the smallest concentration numerically
accessible, 0.1\%, and including $\Ho$ ion concentrations that coincide
with those studied experimentally up to 16.7\%.  Our results suggest
that randomly diluted dipolar Ising systems have a spin-glass phase in
the limit of vanishing dipole concentration, with a critical temperature
vanishing linearly with concentration. The agreement of our results with
mean-field theory testifies to the irrelevance of fluctuations in
interactions strengths, albeit being strong at small concentrations, to
the nature of the low-temperature phase and the functional form of the
critical temperature of dilute anisotropic dipolar systems. Deviations
from linearity in experimental results at the lowest concentrations are
discussed.

\end{abstract}

\date{\today}


\maketitle

\section{Introduction}

Dipolar interactions are ubiquitous in nature and often dominate other
types of interactions in many simple systems, e.g., in insulating
magnets. Additionally, dipolar-like couplings may also arise between
defect states in crystals, where these are mediated by the phonon
vacuum. Such interactions are known to induce magnetic and ferroic order
in densely packed solids and liquids \cite{luttinger:46,vugmeister:90}.
With dilution, the competing nature of the interaction and spatial
disorder lead to a spin-glass (SG) order \cite{binder:86} at low
temperatures.  Mean-field theory suggests that the SG order is
maintained at $x=0^+$, with the critical temperature being linear in the
concentration $x$ \cite{stephen:81,xu:91}.  However, at low
concentrations, spatial inhomogeneities are large and could dominate the
characteristics of the system. Because the dipolar nature of the
interaction renders standard spatial renormalization group methods
ineffective, rigorous analytic conclusions are currently beyond reach.
Thus, the nature of anisotropic dipolar systems in general, and in the
limit of low concentrations in particular, has been a long-standing
controversy.

Experimentally, $\LHx$ is perhaps the best-studied dilute dipolar
(strongly anisotropic) Ising system. This rare-earth compound has
attracted vast experimental, numerical and theoretical interest in the
past two decades. Its scrutinization has enhanced the understanding of
many different magnetic phenomena, such as quantum phase transitions
\cite{bitko:96,ronnow:05,schechter:05}, large spin tunneling
\cite{giraud:01,giraud:03,pollack:14}, quantum annealing
\cite{brooke:99}, quantum entanglement \cite{ghosh:03}, quantum 
domain-wall tunneling \cite{brooke:01}, random-field physics
\cite{schechter:06,tabei:06,schechter:08,silevitch:07} and generic
disordering mechanisms \cite{andresen:13b}. Thus, establishing the
low-temperature phase of $\LHx$ at small $\Ho$ concentrations is not
only of fundamental interest, but is crucial for the further study of
its characteristics. However, extremely long equilibration times have
produced conflicting experimental results and a strong dependence on the
used experimental protocol, with no clear evidence for the equilibrium
phase of the system at low concentrations
\cite{reich:87,ghosh:02,ghosh:03,jonsson:07,silevitch:07b,quilliam:08,quilliam:12,schmidt:14}.
Furthermore, where experimental data suggest the existence of a
spin-glass phase, reported values for the critical temperature deviate
markedly at low concentrations from the expected linear dependence on
the Ho concentration \cite{quilliam:12}.

\begin{figure*}

\includegraphics[width=0.48\textwidth]{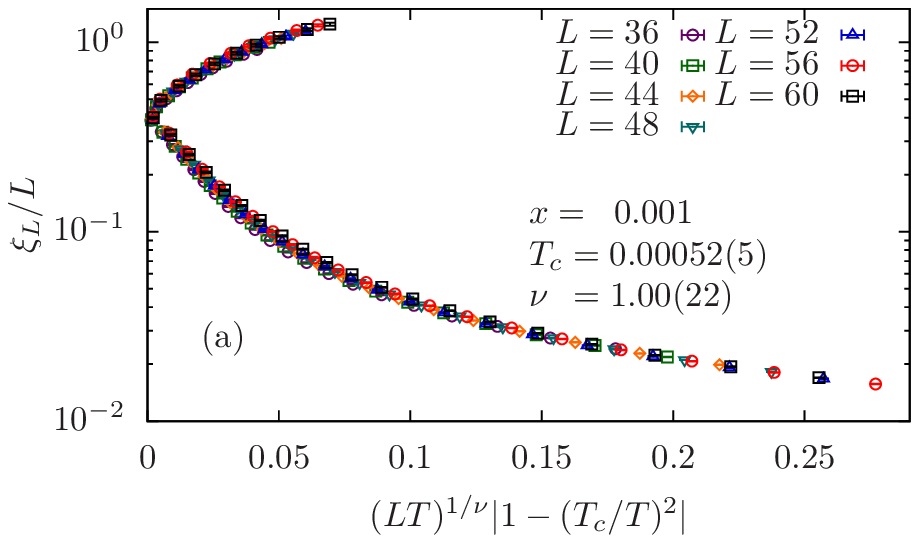}
\includegraphics[width=0.48\textwidth]{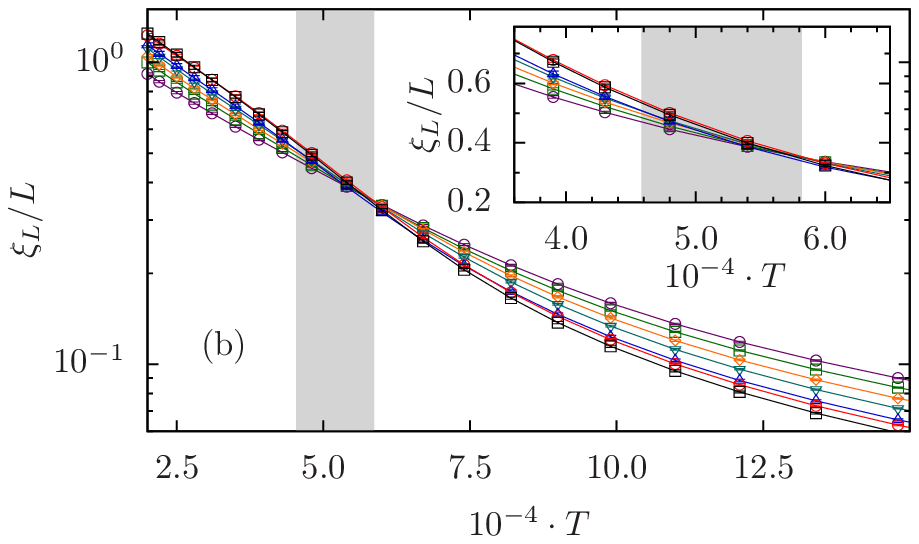}

\caption{
(a) Example finite-size scaling data collapse of the spin-glass
two-point finite-size correlation function $\xi_L/L$ using an extended
scaling form \cite{campbell:06} for the extreme dilution $x=0.001$. The
optimal scaling is accomplished via a Levenberg-Marquardt minimization
procedure described in Ref.~\cite{katzgraber:06}.  (b) Unscaled
two-point finite-size correlation function divided by the linear system
size $\xi/L$ for the extreme dilution $x=0.001$ and system sizes $L=36$,
$\cdots$, $60$.  Data for different system sizes cross at the critical
temperature $T_c$.  The vertical shaded area represents the estimated
critical temperature $T_c=0.00052(5)$K; its width is the statistical
uncertainty determined via an extended finite-size scaling [panel (a)].
The inset zooms into the critical temperature region showing that the
data do indeed cross.
}
\label{fig:collapse}
\end{figure*}

Numerically, understanding the nature of dipolar Ising systems at small
concentrations is notoriously difficult, because spatial inhomogeneities
are large and so are the required system sizes. Previous Monte Carlo
simulations of the dilute dipolar Ising spin-glass model
\cite{ayton:97,snider:05,biltmo:07,biltmo:08} showed no sign of a SG
transition. More recent simulations where a better observable, namely,
the finite-size two-point correlation function \cite{ballesteros:00},
was used suggest the existence of a SG phase down to a concentration of
$x=0.0625$ \cite{tam:09}. However, the regime of much theoretical
interest, where the typical distance between spins is much larger than
interatomic distance, and thus fluctuations are large, could not be
reached. Long equilibration times due to the slow dynamics of the system
\cite{biltmo:12} limited the studied system sizes and concentrations,
i.e., strong finite-size corrections in the data. As such, the nature of
anisotropic dipolar systems at very low concentrations remains unclear.

Here, we present conclusive evidence for the existence of a SG phase in
the dilute dipolar Ising model in the limit of $x=0^+$, and for $\LHx$
for all experimentally relevant low concentrations.  We use large-scale
Monte Carlo simulations that combine parallel tempering
\cite{hukushima:96} and a cluster algorithm \cite{janzen:08,andresen:11}
that allows us to efficiently handle the atypically large interactions
stemming from rare nearby groups of spins and, at the same time, leaves
the prevalent typical interactions for standard numerical treatment. We
find clear evidence that the anisotropic dipolar glass has a SG phase at
low temperatures for all studied concentrations down to $x=10^{-3}$
(almost 2 orders of magnitude smaller than the concentration reached
in previous studies), with a critical temperature $T_c$ that is linear
in the concentration $x$.  Furthermore, our data show that for all $x$,
the divergence of the correlation length at the transition is likely
described by the same critical exponent $\nu$. This strongly suggests
that our results can be carried through to vanishing spin
concentrations.

The paper is structured as follows. In Sec.~\ref{sec:num}, we introduce
a model Hamiltonian for $\LHx$ and outline our numerical approach to
study the system.  We present results in Sec.~\ref{sec:results},
followed by a discussion and concluding remarks in
Sec.~\ref{sec:discussion}.

\section{Model and Numerical Details}
\label{sec:num}

The tunnel splitting induced between the two polarized electronuclear
single Ho ground states in the dilute $\LHx$ system by the crystal field
and by the off-diagonal terms of the dipolar interaction is much smaller
than the typical interaction down to extremely low Ho concentrations
\cite{schechter:08b}. The same is true for the magnetic interaction
between the nuclear spins of the F atoms. Thus, down to very low $x$,
$\LHx$ is well described by a classical Ising spin model
\cite{tam:09,biltmo:12,quilliam:12,andresen:13b}, i.e.,
\begin{align}
\!\!{\mathcal H} \! = \! \sum_{i \ne j}
\frac{J_{ij}}{2} \epsilon_i \epsilon_j S_i S_j
+\frac{J_{\rm ex}}{2} \!\!\sum_{\langle i,j\rangle}
\epsilon_i \epsilon_j S_i S_j \;.
\label{HLiHo}
\end{align}
Here $\epsilon_i=\{ 0,1\}$ is the occupation of the magnetic ${\rm
Ho^{3+}}$ ions on a tetragonal lattice (lattice constants $a = b =
5.175${\AA} and $c = 10.75${\AA}) with four ions per unit cell
\cite{biltmo:09,tam:09}, i.e., $N = 4L^3$ spin sites. $S_i \in \{\pm
1\}$ are Ising spins. The magnetostatic dipolar coupling $J_{ij}$
between two ${\rm Ho^{3+}}$ ions is given by $J_{ij} = D(r^2_{ij} -
3z^2_{ij})/r_{ij}^5$, where $r_{ij}= \vert {\bf r}_{i}-{\bf r}_j\vert$,
${\bf r}_i$ is the position of the $i$-th ${\rm Ho^{3+}}$ ion and
$z_{ij}=( {\bf r}_{i}-{\bf r}_j)\cdot\hat z$ is the component parallel
to the easy axis. The dipolar constant is $D/a^3=0.214$K,
\cite{biltmo:07} and the antiferromagnetic nearest-neighbor exchange is
set to $J_{\rm ex}=0.12$K \cite{biltmo:09}. For the low concentrations
$x$ of interest to us here, this model is equivalent to the pure Ising
dipolar model, because the exchange interactions only slightly change 
the interaction strength of the rare nearby pairs, which, as we show
below, do not affect the thermodynamics at and near the phase
transition.

To determine the finite-temperature transition for a given value of $x$,
we measure the two-point finite-size correlation function
\cite{ballesteros:00}
\begin{equation}
\xi_{L} = \frac{1}{2\sin(k_{\rm min}/2)}\sqrt{
\frac{\av{\aT{q^2(\mathbf{0})}}}
{\av{\aT{q^2(\mathbf{k}_{\rm min})}}}-1}\,,
\label{eq:correlation}
\end{equation}
where
\begin{equation}
q(\mathbf{k}) =
\frac{1}{N}
\sum_{i = 1}^{N} S_i {\rm e}^{i\mathbf{k}\cdot\mathbf{R}_i}\, .
\label{eq:susceptibility}
\end{equation}
Here, $\aT{\cdots}$ represents a thermal average, $\mathbf{R}_i$ is the
spatial location of the spin $S_i$, and $\mathbf{k}_{\rm min}$
represents the smallest nonzero wave vector in the $a$- or $c$-axis
direction, $\mathbf{k}_{\rm min} = (2\pi/L,0,0)$ or $\mathbf{k}_{\rm
min} = (0,0,2\pi/L)$, respectively.  $\xi_L/L$ is dimensionless, and near
the transition, it is expected to scale as $\xi_L/L \sim \tilde
X[L^{1/\nu}(T-T_c)]$. Because corrections to scaling are typically large
for highly dilute systems, we use an extended scaling approach
\cite{campbell:06} that has proven to reduce scaling corrections and
where $\xi_L/L \sim \tilde X[(LT)^{1/\nu}|1-(T/T_c)^2|]$; see
Fig.~\ref{fig:collapse}(a).  When $T=T_c$, the argument of the scaling
function is zero (up to scaling corrections) and hence independent of
$L$.  As such, lines for different system sizes $L$ cross [see
Fig.~\ref{fig:collapse}(b)].  If, however, the lines do not meet, we
know that no transition occurs in the studied temperature range. The
best estimate of the critical temperature $T_c(x)$ is determined by
applying a Levenberg-Marquardt minimization combined with a bootstrap
analysis to determine statistical error bars \cite{katzgraber:06} to the
aforementioned extended finite-size scaling analysis.  An example of the
resulting data collapse using the minimization is shown in
Fig.~\ref{fig:collapse}(a) for $x = 0.001$.

Finally, we note that in Ref.~\cite{tam:09} it was observed that the
estimates of $T_c$ along the $a$ axis tend to be systematically lower
than the ones computed along the $c$ axis of the material.  A comparison
with experimental results \cite{quilliam:12} showed an agreement between
the experimental estimates and the numerical estimates along the $c$
axis only. Similarly, in this work, our estimates of the transition
temperatures from a paramagnetic (PM) phase to a SG phase computed along
the $c$ axis tend to be systematically higher for all studied dilutions
and agree better with the experimental results of Quilliam {\em et
al.}~\cite{quilliam:12}.  As such, all quoted results stem from
simulation results with measurements along the $c$ axis.

In the simulations, we use the Ewald summation method without a
demagnetization factor to compute the periodic boundary conditions
\cite{tam:09,wang:01b} for systems of up to approximately $7\cdot 10^6$
lattice sites. To equilibrate the system at extreme dilutions, we use a
combination of single spin-flip Monte Carlo dynamics and a cluster
renormalization algorithm \cite{janzen:08,andresen:11} combined with
parallel tempering Monte Carlo \cite{hukushima:96}. The cluster
renormalization algorithm is tailored to treat strongly coupled spins
efficiently. It does not fully obey detailed balance; however, it has
been successfully applied to different model systems
\cite{comment:cluster}. The cluster renormalization technique works as
follows: At the beginning of the simulation, set the random positions of
the spins and search for clusters $C_{J_i}^i$ of spins coupled by an
interaction of at least $\vert J_i\vert$. Once all clusters have been
labeled and recorded, search for sets of spins $C_{J_{i+1}}^{i+1}$
coupled by at least $\vert J_{i+1}\vert=\vert J_{i}\vert +\delta_J$ and
record these.  Iteratively perform this renormalization $i = 0$, $1$,
\ldots, $n$ times until all the sets of $C_{J_n}^n$ consist of only spin
pairs.  It is very important to carefully tune the renormalization
procedure to the studied model. In this case, we use $\vert J_0\vert=x$,
where $x$ is the concentration of the magnetic ions in the system. A
suitable step value $\delta_J$ is $2x$. Spins in clusters are then
flipped, regardless of their sign. Note that one Monte Carlo sweep in
the simulation thus consists of the following procedure: For each spin
in the system, we either perform a single-spin simple Monte Carlo flip
with probability $0.75$, or we flip a randomly selected cluster.

\begin{table}[h]
\caption{
Simulation parameters for the different concentrations $x$ and linear
system sizes $L$ studied.  $T_{\rm min}$ [$T_{\rm max}$] is the lowest
[largest] simulated temperature, and $N_{\rm T}$ is the number of
temperatures used in the parallel tempering Monte Carlo method.  We
thermalize and measure for $2^X$ Monte Carlo sweeps, and $N_{\rm sa}$ is
the number of disorder realizations.  Note that to compute the
spin-glass order parameter, we actually simulate $2N_{\rm T}$ replicas
($N_{\rm T}$ for the tempering method and two per temperature to compute
the spin overlap). Note that smaller system sizes have also been
studied; however, for brevity, we do not list them because they share the
same parameters as the smallest $L$ listed for a given concentration
$x$.
\label{tab:simparams}}
{\scriptsize
\begin{tabular*}{\columnwidth}{@{\extracolsep{\fill}} l r r r r r r}
\hline
\hline
$x$     & $L$ & $T_{\rm min}$ & $T_{\rm max}$ & $N_{\rm T}$ & $X$ & $N_{\rm
sa}$ \\
\hline
$0.001$  & $36$ & $0.0002$   & $0.005$ & $30$ & $17$  & $2500$ \\
$0.001$  & $40$ & $0.0002$   & $0.005$ & $30$ & $19$  & $3235$ \\
$0.001$  & $44$ & $0.0002$   & $0.005$ & $30$ & $20$  & $1493$ \\
$0.001$  & $48$ & $0.0002$   & $0.005$ & $30$ & $21$  & $1486$ \\
$0.001$  & $52$ & $0.0002$   & $0.005$ & $30$ & $22$  & $1092$ \\
$0.001$  & $56$ & $0.0002$   & $0.005$ & $30$ & $22$  & $1016$ \\
$0.001$  & $60$ & $0.0002$   & $0.005$ & $30$ & $22$  & $640$ \\[1mm]
$0.018$  & $20$ & $0.0069$   & $0.065$ & $22$ & $19$  & $1270$ \\
$0.018$  & $22$ & $0.0069$   & $0.065$ & $22$ & $20$  & $1274$ \\
$0.018$  & $24$ & $0.0069$   & $0.065$ & $22$ & $20$  & $1019$ \\[1mm]
$0.045$  & $14$ & $0.0151$   & $0.200$ & $27$ & $19$  & $1983$ \\
$0.045$  & $16$ & $0.0151$   & $0.200$ & $27$ & $19$  & $1040$ \\
$0.045$  & $18$ & $0.0151$   & $0.200$ & $27$ & $20$  & $1037$ \\[1mm]
$0.080$  & $12$ & $0.0356$   & $0.200$ & $28$ & $18$  & $1981$ \\
$0.080$  & $14$ & $0.0356$   & $0.200$ & $28$ & $19$  & $1987$ \\
$0.080$  & $16$ & $0.0356$   & $0.200$ & $28$ & $19$  & $0704$ \\[1mm]
$0.167$  & $ 8$ & $0.0712$   & $0.300$ & $25$ & $16$  & $1488$ \\
$0.167$  & $10$ & $0.0712$   & $0.300$ & $25$ & $17$  & $1000$ \\
$0.167$  & $12$ & $0.0839$   & $0.300$ & $23$ & $18$  & $1020$ \\[1mm]
$0.198$  & $ 8$ & $0.1100$   & $0.320$ & $40$ & $15$  & $1980$ \\
$0.198$  & $10$ & $0.1100$   & $0.320$ & $40$ & $16$  & $1012$ \\
$0.198$  & $12$ & $0.1100$   & $0.320$ & $40$ & $18$  & $1105$ \\[1mm]
$0.250$  & $ 8$ & $0.1800$   & $0.450$ & $25$ & $14$  & $2800$ \\
$0.250$  & $10$ & $0.1800$   & $0.450$ & $25$ & $14$  & $1054$ \\
$0.250$  & $12$ & $0.1800$   & $0.450$ & $25$ & $17$  & $1049$ \\
\hline
\hline
\end{tabular*}
}
\end{table}

Finally, to verify that the data are properly thermalized, a logarithmic
binning analysis is used: Observables are measured and averaged over an
exponentially growing number of Monte Carlos sweeps $2^X$, and their
Monte Carlo time evolution is monitored. When at least three bins
agree within error bars and are independent of Monte Carlo time, we deem
the system to be in thermal equilibrium. Simulation parameters are
listed in Table \ref{tab:simparams}.

\begin{figure}
\includegraphics[width=3in]{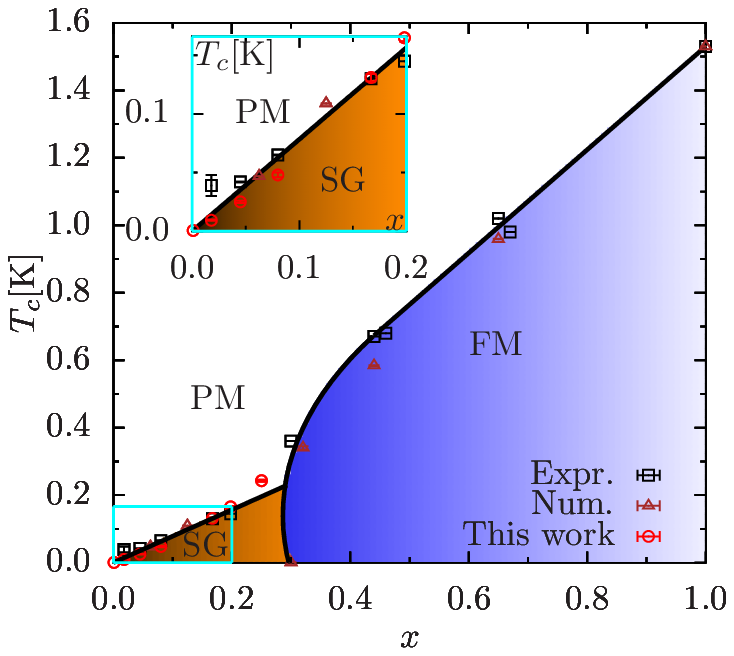}
\caption{
Transition temperature $T_c$ versus concentration $x$ phase diagram
obtained from our simulations (circles), previous simulations from
Refs.~\cite{tam:09,andresen:13b} (triangles) and experimental data from
Refs.~\cite{kjaer:89,bitko:96,silevitch:07,ancona-torres:08,quilliam:12}
(squares).  The straight line between the spin-glass (SG) and
paramagnetic (PM) phases is a linear fit to the data for small
concentrations.  Our results suggest that the spin-glass phase extends
to the $x=0^+$ limit, as can be seen in more detail in the inset.  For
large concentrations and low temperatures, the system is a ferromagnet
(FM).
}
\label{fig:phase_diagram}
\end{figure}

\begin{figure}
\includegraphics[width=3in]{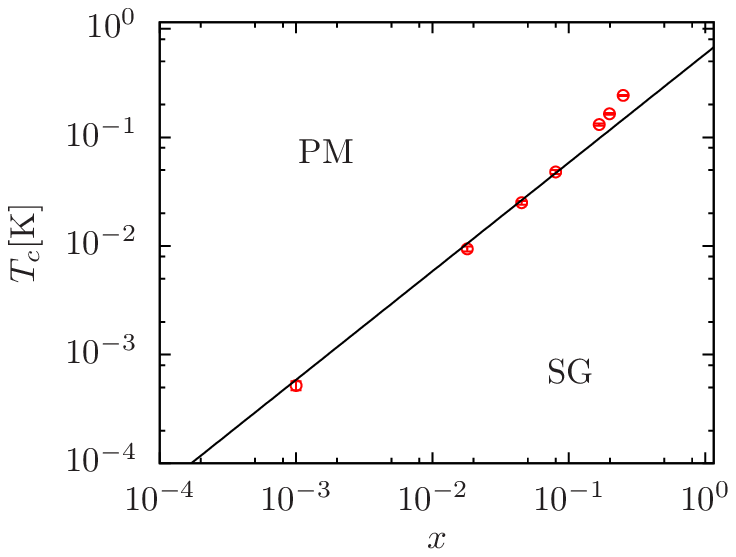}
\caption{
Log-log plot of the high-dilution limit of the transition temperature
$T_c$ versus the concentration $x$ phase diagram obtained from our
simulations.  The solid line separating the spin glass (SG) from the
paramagnetic (PM) phase is a fit to the numerical data of the form
$T_c(x) = ax$ with $a=0.59(1)$. The spin-glass phase therefore extends
to the  $x=0^+$ limit.
}
\label{fig:fit}
\end{figure}

\begin{figure}
\includegraphics[width=3in]{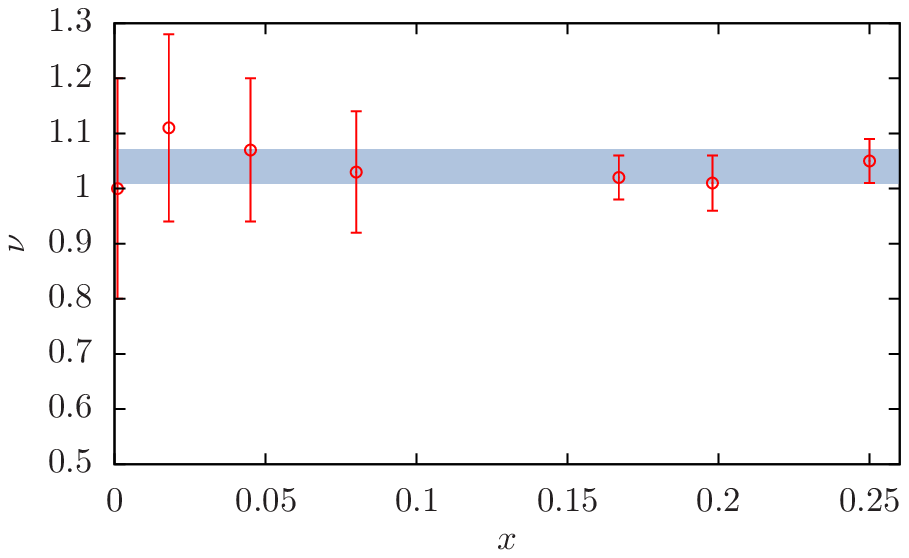}
\caption{
Critical exponent $\nu$ as a function of the concentration $x$.  The
shaded area corresponds to their average value over all concentrations.
The individual estimates come from an extended finite-size scaling
analysis of the two-point correlation length; an example is shown in
Fig.~\ref{fig:collapse}(a).  All estimates agree within error bars,
meaning that the critical exponent $\nu$ might be independent of the
concentration $x$. This hints towards the possibility of a common
universality class.
}
\label{fig:universality}
\end{figure}

\section{Results}
\label{sec:results}

Our main result is shown in the phase diagram depicted in
Fig.~\ref{fig:phase_diagram}, as well as Table \ref{tab:results}. The
estimated critical temperatures $T_c(x)$ show a clear linear behavior
for 2 orders of magnitude, strongly suggesting that the SG phase extends
to the zero concentration limit.  Comparison to experiment shows that
for $x=0.167$, we find $T_c=0.131(3)$, in excellent agreement with the
experimental results of $T_c=0.133(5)$ \cite{wu:91}. Similarly, for
$x=0.08$, we find $T_c=0.048(2)$, close to the most recent experimental
result, $T_c=0.065(3)$ \cite{quilliam:12}. For lower concentrations,
however, we find values of $T_c$ which agree well with the linear
extrapolation of the experimental data from higher $x$ but are lower
than the experimentally obtained values for $x=0.045$ and $x=0.018$.  We
attribute this discrepancy to the microscopic time scale in $\LHx$ being
very long at low temperatures and enhanced with the decrease of the Ho
concentration. This results in the difficulty to equilibrate the system
close to the critical temperature at the lowest experimentally studied
concentrations \cite{quilliam:12,schmidt:14}. Furthermore, it was argued
that long equilibration times of small clusters lead to a quantum
nonequilibrium state, whose nature depends on the degree of coupling to
the environment \cite{schmidt:14}.  

In Fig.~\ref{fig:fit}, we show a log-log plot of the high-dilution limit
of the transition temperature $T_c$ versus concentration $x$ phase
diagram shown in the inset to Fig.~\ref{fig:phase_diagram}. The solid
line in the figure that separates the SG from the PM phase is a fit to
$T_c(x) = ax$ with $a=0.59(1)$.  Allowing for a finite intercept, i.e.,
$T_c(x) = T_c^0 + ax$, yields $a = 0.60(2)$ and $T_c^0 = -0.00097(94)$,
which is statistically compatible with a zero intercept.  Therefore, we
see strong evidence that the spin-glass phase extends to the $x=0^+$
limit.

We note that despite the much-enhanced fluctuations in the distribution
of interactions as $x$ is reduced, equilibration times are similar for
all system sizes. The simulations for the lowest concentration were
limited by the time it takes to accurately compute the Ewald summation
used to account for the periodic boundary conditions, and not the Monte
Carlo simulation time, therefore showing the effectiveness of the
implemented algorithm even for very high dilutions.  This bottleneck is
in fact much easier to overcome because the Ewald summation has to be
performed only once for each system size at the beginning of the
simulation.

\begin{table}
\caption{
Critical temperature $T_c$ and critical exponent $\nu$
extracted from an extended scaling analysis for all studied
concentrations $x$.
\label{tab:results}}
{\scriptsize
\begin{tabular*}{\columnwidth}{@{\extracolsep{\fill}} c l l}
\hline
\hline
$x$     & $T_c$ & $\nu$\\
\hline
$0.001$  & $0.00052(5)$ & $1.00(20)$   \\
$0.018$  & $0.0094(5)$  & $1.11(17)$   \\
$0.045$  & $0.025(1)$   & $1.07(13)$   \\
$0.080$  & $0.048(2)$   & $1.03(12)$   \\
$0.167$  & $0.131(3)$   & $1.02(4)$    \\
$0.198$  & $0.165(3)$   & $1.01(5)$    \\
$0.250$  & $0.243(3)$   & $1.05(4)$    \\
\hline
\hline
\end{tabular*}
}
\end{table}

In Fig.~\ref{fig:universality}, we show the critical exponent $\nu$ as a
function of $x$ computed from a finite-size scaling of the two-point
correlation function. For all concentrations depicted, data seem to
agree approximately within error bars---an indicator for potential
universal behavior. All critical parameters from the scaling analysis of
the two-point correlation function are listed in Table
\ref{tab:results}.  Note, however, that to fully characterize a
universality class, it is necessary to determine at least two critical
exponents.  We tried different analysis methods to extract a second
critical exponent $\eta$ from the spin-glass susceptibility.  However,
no robust estimate was possible, a common problem in spin-glass
simulations (see also Ref.~\cite{katzgraber:06}).

\section{Discussion}
\label{sec:discussion}

Dilute power-law interacting systems are natural candidates for emergent
geometric similarity \cite{cederberg:01} whereby statistical mechanics
of systems at different concentrations may be mapped onto each
other~\cite{larkin:70}. The characteristic $1/r^d$ falloff of the
dipolar kernel implies linear scaling of typical interactions with
concentration in any dimension $d$ and suggests similar scaling of
relevant temperature scales.  However, dipolar systems at different
concentrations are not quite geometrically similar. Rescaling of the
interactions leaves the distribution of interactions practically
unchanged at low and typical values but generates a progressively
stronger tail at high values, because the largest coupling is fixed by
the lattice spacing, independent of the concentration.  These large
couplings produce physical correlation effects  that impeded simulation
progress in the past.  Remarkably, focusing on and solving this
relatively local high-energy bottleneck allows for essentially unimpeded
progress on the rest of the problem.  Our results provide strong support
to the notion of emergent geometric similarity by locating and
characterizing the spin-glass ordering transition over nearly 2 orders
of magnitude in concentration, with the transition temperature scaling
linearly with concentration.

We note here that this geometric similarity is destroyed by the
application of a transverse field \cite{schechter:09}, which, in
combination with the off-diagonal elements of the dipolar interaction,
results in effective random fields in the longitudinal direction
\cite{schechter:06,tabei:06,schechter:08}. The emergent longitudinal
fields have a large variance, are correlated with the interactions, and
lead to a much more effective disordering of the spin-glass phase
\cite{schechter:09} than that predicted by the naive application of the
Imry-Ma argument \cite{imry:75,fisher:86}. We emphasize that in the
absence of an applied field, the effect of the off-diagonal dipolar terms
on the thermodynamic phase of the system is negligible. The quantum
fluctuations induced by these is much smaller than the interaction, and
they do not change the ${\mathbb Z}_2$ time-reversal symmetry (spin
inversion) of the Hamiltonian.

Summarizing, using large-scale Monte Carlo simulations that combine
parallel tempering with an innovative cluster renormalization algorithm
\cite{janzen:08,andresen:11}, we have shown that the dilute dipolar
Ising model has a spin-glass transition at low temperatures for
concentrations down to $x = 10^{-3}$. Furthermore, a clear linear
behavior of $T_c(x)\sim x$ is found in the highly dilute regime,
strongly suggesting that the SG phase transition extends to the $x=0^+$
limit.

\begin{acknowledgments}

We are grateful to A.~Aharony, B.~Barbara, M.~Gingras, S.~Girvin,
D.~Huse, J.~Kycia, D.~Sherrington, and D.~Silevitch for enlightening
discussions.  H.G.K. acknowledges support from the NSF (Grant No.
DMR-1151387), thanks Hitachino Nest for inspiration, and would like to
thank the Santa Fe Institute for their hospitality. V.O.~acknowledges
support from the NSF (Grant No.~DMR-0955714).  M.S.~acknowledges support
from the Marie Curie Grant No. PIRG-GA-2009-256313 and from the ISF
Grant No.~821/14.  We thank the Texas Advanced Computing Center (TACC)
at The University of Texas at Austin for providing HPC resources
(Stampede Cluster), ETH Zurich for CPU time on the Brutus cluster, and
Texas A\&M University for access to their Eos and Lonestar clusters.

\end{acknowledgments}

\bibliography{refs,comments}

\end{document}